\documentstyle[aps,preprint]{revtex}

\begin{document}

\title{Landscape Paradigms in Physics and Biology: Introduction and Overview}
\author{David Sherrington}
\address{Dept. of Physics, Oxford University, Theoretical Physics, 1 Keble Rd., Oxford OX1 3NP, England
\linebreak  Center for Nonlinear Studies, Los Alamos National Laboratory, Los Alamos, NM 87545, USA}
\maketitle

\begin{center}
July 8, 1996
\end{center}

\begin{abstract}
A brief introductory overview in general terms is given of concepts, issues  and applications of the paradigm of rugged landscapes in the contexts of physics and biology.
\end{abstract}

In the present context, landscapes describe the structure of control functions relevant to the cooperative behaviour of systems of many interacting units. The paradigm is now ubiquitous in several branches of science, particularly for the conceptualization of behaviour which is commonly described as complex.

To biologists  the landscape is typically visualized as giving a measure of fitness, to be maximized, while to physicists it is usually considered as specifying  an energy, to be minimized. However, these are simply inverted representations  of the  of the same thing and henceforth I shall tend to use the physicists' language. 

As in geography, these landscapes come in various forms, flat, smooth, discontinuous  and rugged. Again as in human perception of the world around us, flat landscapes are the simplest to contemplate but rugged landscapes excite the greatest interest, in the sense of having the richest, most complex consequences. This perception of interest in ruggedness is reflected in the papers that follow, but even systems with flat landscapes can yield  highly non-trivial behaviour when interactions between their inhabitants are sufficiently complicated. 

What is the space of these landscapes? For some problems it is low-dimensional, as,for example, when the landscape represents the potential energy seen by an electron due to interactions with atoms, ions or other fixed objects; in this case the `grid coordinates' (to use an analogy with a cartographic map) indicate the location of the electron and the `elevation' measures the potential at that point \cite{Mackinnon}. A similar situation at a more coarse-grained level can apply to measures of coefficients in a Ginzburg-Landau free energy functional expansion in statistical physics. Here, however, we shall be thinking of a different situation where the landscape sits in some high-dimensional space in which each `grid-point' specifies either a complete microstate, which describes the `positions' of {\em all} the individual units which make up the many-body system, or some more coarse-grained, but still multi-dimensional, macrostate characterization. In most cases we shall think in terms of a single `height' parameter as a function of a multi-dimensional `location' parameter, but it is perfectly possible to have a several-dimensional height measure. Neither the `horizontal' nor the `vertical' coordinates of the landscape need be continuous, but, since the conventional world whose experience has molded our normal conceptualization does have this feature, for orientation I shall often use images based on such a continuous picture. 

At the simplest level one might think of the dynamical behaviour of the many-body system in terms of motions on this landscape; in particular, for deterministic dynamics, in terms of gradient descent to local minima. In a non-flat landscape this leads immediately to an image of separated regions of flow and their associated attractors, corresponding to the valleys, with barriers between them, corresponding to the hills and saddles. 

For many of the problems of interest the effective landscape structure is rugged in the sense that within a single closed contour of some `height' one finds many closed contours of lower height and, further, within each of these one finds many closed contours of even lower height and so on. Equivalently, there is a hierarchy of  several sub-valleys within valleys at many scales. The consequence is that deterministic microdynamics can yield many possible final stopping states, hierarchically related and often with quasidegeneracies in their `heights'. Even for stochastic dynamics, in which uphill moves are also allowed with a probability decreasing with the height change involved, in large enough systems ruggedness of the microscopic landscape leads to the possibility of effective non-ergodicity in which non-equivalent macrostates result depending on the starting microstates, not communicating on realistic timescales. This is sometimes re-expressible as downhill  moves on a still-rugged {\em free energy} surface in macrospace.   

Let us turn now to the origin of rugged landscapes. They can arise due to competition between different microscopic few-body interactions; for example, in a magnetic context  between ferromagnetic and antiferromagnetic exchange interactions; in a neural network between excitatory and inhibitory synapses or neurons. Or they can be due to conflicts between few body-forces and global constraints, as in the cost functions of graph equipartioning. They can also have their origin in competition between internal and external forces, as in the random field Ising model. Generically we refer to these conflicts as {\em frustration}.

On a fine scale the contours of a landscape can vary slowly and continuously or they can involve a series of quasi-steps at which the height changes rather rapidly, separating regions of slower but still hierarchical evolution; in the first case one would expect a continuous hierarchy of metastable states as in some spin-glass models \cite{Mezard} , 
while the second would suggest `tiers' of `conformational substates' as suggested in some studies of proteins (like myoglobin) \cite{Frauenfelder}. 

The landscapes themselves are not immutable but can change with changes in interactions or external perturbations. Such changes can be (quasi-)continuous
on some longer timescale, such as in long time potentiation (or synaptic modification) in neural network learning, or they can be sudden due to a fast perturbation, such as occurs in photoexcitation in proteins. In principle, some of these modifications can be considered within a larger space of dynamics allowing for simultaneous evolution of both the landscape and the elements it controls, typically with different timescales. Changes can be smooth or chaotic in their response to changes in global control parameters.

Another distinction we should make is between random and quasi-random ruggedness. In some systems, the landscape is controlled by truly random quantities; an example is a spin glass model in which the  the exchange interactions are randomly chosen and are thought to give rise to an energy landscape with very many attractors, a high quasi-degeneracy of the energies of these attractors and slow long-time dynamics to approach them. In other systems, however, the landscape is sculpted via appropriate changes in the controlling few-body interactions. One example is in a neural network trained via a supervised learning procedure to yield desired (memory) attractors. More instances of rugged but less truly random landscapes arise in several biological, economic and ecological contexts tuned for success; for example, a biologically relevant (and realistic) protein must have dynamics leading quickly to a folded state with appropriate structure and function, suggesting that it should have one large and dominant attractor, possibly with several quasi-degenerate `ground states' with similar functions but with large energy separations from  higher states with different functions \cite{Wolynes,Analogy}; while in much of nature the successful agents are those which have evolved to perform their tasks efficiently and robustly. Henceforth I shall refer to this second group as `sculpted landscapes' \cite{quasi-random}.

Thus far we have considered only the nature of the landscapes. However,  the allowed `steps' are also very important and the attractor structure and dynamics can be different in different `step-spaces'. Furthermore, the whole concept of motion via descents on a landscape is itself only a special case of a more general dynamic flow space for which it is not necessary to have detailed balance or a description in terms of a quantity which is always minimized in each microscopic move (a Lyapunov function); as examples of such systems one can quote some neural networks and most cellular automaton models. More fundamental is the structure of the space of dynamical flows, which is often hierarchically fractured (and therefore warrants a description as `rugged') without having a true landscape description; it is clear that if one is to have non-fixed point attractors, such as sequences or restricted strange attractors, the naive landscape description needs such extension. We shall take the landscape paradigm to be generalized in this sense. However, for convenience we shall continue to use the language of the naive landscape paradigm in general discussion.

As noted earlier, motion on the landscapes need not not be deterministic, but can also be stochastic; the moves need not be simply locally downhill in the space of energy landscapes over the full microspace, but one can have `probabalistic hill-climbing'; alternatively, in the description employing free energies in macrospace one has a modification of the landscape itself, typically a smoothing with increased temperature, but also with possibilities of entropically driven new attractors. A similar situation applies to the modification of flows, attractors and their basins  even without detailed balance. These changes can lead to phase transitions as a function of stochasticity as well as those due to changes in global control parameters. This is also an appropriate point to emphasise the difference between thermodynamic and attractor phase diagrams; the former are concerned with systems with detailed balance dynamics, governed by the laws of equilibrium statistical mechanics and with states weighted by Bolzmann or relevant quantum statistical factors, while the latter is concerned with the occurence of dynamical attractors, even in systems without Lyapunov functions and where, even if the concept is meaningful, their energies may be so high as to exclude them from thermodynamic relevance.

There are clearly many different systems which can be considered in the terms discussed so far. Some require quite different mathematical formulation but other physically quite distinct systems can be described mathematically in very similar fashions. As an illustration let us consider a set I have referred to previously \cite{Sherrington} under the grouping ``Magnets, microchips and memories'', in which the magnets are Ising spin glasses, the microchips refer to the problem of equally bipartitioning the elements of an electrical circuit between two microchips so as to minimize the number of wires between the chips, and the memories refer to recurrent neural networks of McCulloch-Pitts neurons. All these cases can be described by an energy function  $H = - \sum_{(ij)} J_{ij} \sigma_{i} \sigma_{j}$ where the subscripts label the microscopic units ( spins, circuit elements, neurons); respectively for the magnets, microchips and memories, $\sigma = +/-1$ refers to whether the spins are up/down, the circuit elements are on the first/second microchip, or the neurons are firing/nonfiring, while the $J_{ij}$ measure exchange interactions between spins, the wirelength needed if two circuit elements are on different microchips, or the strength and character of the synaptic influence of one neuron on another. For the magnets and memories the $J$'s are a mixture of positive and negative signs, while for the microchips the $J$'s are all positive but there is a global constraint that $\sum_{i} \sigma_{i} =0$. 
All these situations are frustrated, yielding rugged landscapes. The $J$'s are typically random for the magnets, only quasi-random for the neural networks which are sculpted to yield attractors corresponding to memorized global patterns, and quasi-random, but not sculpted, for the microchips if the circuit connections are simply designed to yield an appropriate electronic operation rather than to optimize the placement and wiring problem discussed here. Furthermore, both the magnets and the memories can be  considered also within a related context in which only a dynamic description is given, that the probability of updating $\sigma_{i}$ is determined only by the instantaneous value of a `field' $h_{i}=\sum_{j}J_{ij}\sigma_{j}$. In the case of the neurons  there is no need for $J_{ij}$ to be equal to $J_{ji}$ and hence there need be no `energy' landscape.  For the microchip problem there is no {\em a priori} dynamics; the objective is to find a dynamics which minimizes the cost and use it to find the corresponding microstate.

As noted earlier, the dynamical behaviour of random and sculpted landscapes are typically quite different. For random systems the long time dynamics is usually slow, although there may be faster initial transients, whereas for survival in the world the kinetics of achieving a desired state are as important as the latter's structure and require an appropriate tuning of the landscape; for example, prey must respond quickly to the presence of predators, a protein must fold rapidly, and a neural network must associate or generalize quickly.

Usually one does not require knowledge of the full micro-description, but rather one wants macroscopic measures to monitor performance. Hence it is important to consider the passage from the full microdynamics to a consequential macrodynamics in terms of a few macro-observables. These macroparameters can be a set of instantaneous measures or they can involve multi-time correlation and response functions. Whatever the specific case, one is faced with the question of how many such macrovariables are needed for an adequate description; typically one cannot express the dynamics of disordered and frustrated systems adequately in terms of a single instantaneous macroparameter. Studies of the macrodynamics expose such features as (slow) glassy macrodynamics and aging in systems with very rugged landscapes.

Also relevant at the level of macrodynamics is the question of self-averaging, or its absence. Self-averaging refers to the feature of independence of macro-observables from the specific instances of microscopic randomness, with only the distributions from which the random elements are drawn being relevant in the limit of large systems.  Dependent on the system, some macrovariables are self-averaging while others are not. For example, in infinite-ranged spin glasses the energy and the magnetization are self-averaging with respect to the specific choice of random exchange interactions, but the overlaps \cite{overlap} between two identical but separately evolving replicas are not. There are also typically sample-to-sample fluctuations among the reaction rates of folded proteins with the same molecular sequences. For this reason care is required in specifying different types of averages. A related concept is that of ultrametricity \cite{ultrametricity,Mezard} of such non-selfaveraging quantities, itself an indication of the hierarchical organization of the determining landscape.

The evidence for the images presented above comes from a combination of `real' experiments, computer simulation experiments and the analysis of theoretical models. In the papers which follow several aspects of both evidence and consequences will be discussed in many different systems with many different investigative techniques. The fundamental issues can be boiled down to two questions; (i) at the specific system level, how do the microscopics lead to macroscopic structure and function and/or what do the observed structure and function tell about the microdynamics, (ii) at the global level, to what extent is there universality in the landscape paradigm in different areas of science and compementarily what are the nuances of differences? The simplicity of these questions hides the very considerable subtlety of their answers and of the further questions they raise. They have exercised many brains for several decades before yielding the conceptual images and the experimental, analytic and computational techniques which are now in place and which have greatly enriched our understanding of the complex world around us and our toolbox of ways to probe it, yet the knowledge we have gained so far is certainly only the tip of an iceberg whose true majesty will take many more years to be exposed.

{\bf Acknowledgements}
My own understanding of the landscape paradigm owes much to discussion with many colleagues and friends. They are too numerous to name here but they know who they are and I do thank them sincerely.


\begin{thebibliography}{ultrametricity}

\bibitem{Mackinnon} The physics of such systems is encapsulated in the title of Angus Mackinnon's inaugural lecture as a Professor at Imperial College, ``Bens, glens, lochs, and burns'', which  translates from Scottish to English as ``Mountains, valleys. lakes and streams'', the first two nouns characterizing the landscape and the last two its consequences.

\bibitem{Mezard} See for example M.M\'ezard, G.Parisi and M.Virasoro, {\it Spin Glass Theory and Beyond} (World Scientific, Singapore 1987).


\bibitem{Frauenfelder} For a summary see H.Frauenfelder, S.G.Sligar and P.G.Wolynes, Science {\bf 254}, 1598 (1991) 

\bibitem{Analogy} I might mention that flow towards a solution having unique function with respect to a particular measure but varying microdetails with respect to others is not uncommon; for example, in the neural memory task of person recognition one regularly identifies and responds emotionally to another person independently of their facial expression, clothes etc., even when the latter are noticed.

\bibitem{Wolynes} Hence the picture of a `folding funnel'; see for example P.G.Wolynes, J.N.Onucic and D.Thirumalai, Science  {\bf 267}, 1619 (1995).

\bibitem{quasi-random} Although some of these systems might seem random to a casual observer this `randomness' is analagous to that of a language of which one is ignorant; like the Navajo `code' used by the Americans in the last world war and unbroken by the Japanese, or like Japanese script to an American (Navajo or Caucasian) who does not know how to read it.

\bibitem{Sherrington} D.Sherrington, Spec.Sci.Tech. {\bf 14}, 316 (1991)

\bibitem{overlap} The overlap of two states is a measure of their microscopic similarity; for a system of binary microvariables it is given by the fraction of microvariables which are identical in the two microstates minus the fraction which are different.

\bibitem{ultrametricity} Given three states $S,S',S''$, their overlaps $q(S,S')$, $q(S',S'')$ and $q(S'',S)$ are ultrametrically related if the two smallest overlaps are equal.

\end{thebibliography}
\end{document}